\documentstyle[preprint,aps,12pt,epsf]{revtex}
\tightenlines
\input epsf
\begin{document}
\draft
\preprint{\vtop{{\hbox{YITP-00-49}\vskip-0pt
}}}
\title{Contribution of sigma meson pole \\
to $K_L$- $K_S$ mass difference
\footnote{Talk given at the Workshop on Possible Existence of the 
$\sigma$-meson and Its Implications to Hadron Physics, June 12 -- 14,
2000, Yukawa Institute for Theoretical Physics, Kyoto University, 
Kyoto}} 
\author{K. Terasaki\\ Yukawa Institute for Theoretical Physics,\\
Kyoto University, Kyoto 606-8502, Japan
}
\maketitle
\thispagestyle{empty}
\begin{abstract}
The hypothesis of $\sigma$ meson pole dominance in the 
$|\Delta {\bf I}|={1\over 2}$ $K\rightarrow \pi\pi$ amplitudes 
is tested qualitatively by using the $K_L$-$K_S$ mass difference. 
\end{abstract}

\vskip 0.5cm

Dominance of $\sigma$-meson pole contribution in the amplitudes for 
the $K_S\rightarrow \pi\pi$ decays was first proposed as the origin 
of the well-known $|\Delta {\bf I}|={1\over 2}$ rule in these 
decays~\cite{rf:sigma}, and recently revived in connection with the 
direct $CP$ violation in the $K\rightarrow \pi\pi$
decays~\cite{rf:CP}. If it is the case, however, the matrix elements, 
$\langle{\sigma|H_w|K}\rangle$, should survive and give a significant 
contribution to the $K_L$-$K_S$ mass difference, $\Delta m_K$, where 
$H_w$ is the strangeness changing ($|\Delta S| = 1$) effective weak 
Hamiltonian. 

Dynamical contributions of various hadron states to hadronic 
processes in which pion(s) take part can be estimated by using a hard 
pion technique (with PCAC) in the infinite momentum frame 
(IMF)~\cite{rf:hard-pion}. 
For later convenience, we review briefly it below. As an example, we 
consider a decay, $B(p) \rightarrow \pi_1(q)\pi_2(p')$, in the IMF,
i.e., ${\bf p}\rightarrow \infty$, and assume that its amplitude 
$M(B \rightarrow \pi_1\pi_2)$ can be approximately evaluated at a
slightly unphysical point, ${\bf q} \rightarrow 0$, i.e., 
$q^2\rightarrow 0$ but $(p\cdot q)$ is finite:
\begin{equation}
 M(B\rightarrow \pi_1\pi_2) 
 \simeq \lim_{{\bf p} \to \infty,\,\, {\bf q} \to 0}
M(B\rightarrow \pi_1\pi_2) .
\end{equation}
In this approximation, the $\sigma\rightarrow\pi^+\pi^-$ amplitude 
is described in terms of the {\em asymptotic} matrix element, 
$\langle{\pi^-|A_{\pi^-}|\sigma}\rangle$, (matrix element of 
$A_{\pi^-}$ taken between $\pi^-$ and $\sigma$ with infinite 
momentum) as 
\begin{equation}
M(\sigma\rightarrow \pi^+\pi^-) 
\simeq \sqrt{2}\Bigl({m_{\sigma}^2 - m_{\pi}^2 
                                 \over f_{\pi}}\Bigr)
  \langle{\pi^-|A_{\pi^-}|\sigma}\rangle , 
                                             \label{eq:sigma-pipi}
\end{equation}
which has been symmetrized with respect to exchange of $\pi^+$ and 
$\pi^-$ in the final state since isospin symmetry is always assumed 
in this note. The asymptotic matrix element, 
$\langle{\pi^-|A_{\pi^-}|\sigma}\rangle$, is given by 
\begin{eqnarray}
&& \hspace{-10mm}
\lim_{{\bf p}\rightarrow\infty,\,\,{\bf q} \rightarrow 0}  
\langle{\pi^-(p')|A_{\pi^-}|\sigma(p)}\rangle    \nonumber\\
&& =(2\pi)^3\delta^{(3)}({\bf p} - {\bf p'})
\langle{\pi^-|A_{\pi^-}|\sigma}\rangle
\sqrt{N_{\pi}N_\sigma}\Big|
_{{\bf p}={\bf p'}\rightarrow\infty}
\end{eqnarray}
and is related to the $\sigma\pi\pi$ coupling constant in the usual 
Feynman diagram approach~\cite{rf:hard-pion}, where $N$ is the 
normalization factor of state vector. 

Using the same technique, we can describe dynamical contributions of 
hadrons to the $K \rightarrow \pi\pi$ amplitude by a sum of 
equal-time commutator (ETC) term and surface term, 
\begin{equation}
 M(K \rightarrow \pi_1\pi_2) 
\simeq M_{\rm ETC}(K\rightarrow \pi_1\pi_2) 
+ M_{\rm S}(K\rightarrow \pi_1\pi_2).            \label{eq:hard pion}
\end{equation}
$M_{\rm ETC}$ has the same form as that in the old soft pion 
technique~\cite{rf:soft-pion}
\begin{equation}
M_{\rm ETC}(K\rightarrow \pi_1\pi_2)
= {i \over \sqrt{2}f_{\pi}}
     \langle{\pi_2|[V_{\bar \pi_1}, H_w]|K}\rangle 
                      + (\pi_1 \leftrightarrow \pi_2)   \label{eq:ETC}
\end{equation}
but it now should be evaluated in the IMF. The surface term, 
\begin{equation}
M_S(K\rightarrow \pi_1\pi_2) 
=  \lim_{{\bf p} \to \infty,\,\,{\bf q} \to 0}
\Bigl\{-{i \over \sqrt{2}f_\pi}q^\mu T_\mu\Bigr\}
+ (\pi_1 \leftrightarrow \pi_2), 
\end{equation}
survives in contrast with the soft pion approximation 
and is now given by a sum of all possible pole amplitudes, 
\begin{equation}
M_{\rm S} = \sum_n M_{\rm S}^{(n)} + \sum_l M_{\rm S}^{(l)}, 
                                               \label{eq:surf-tot}
\end{equation}
where the hypothetical amplitude $T_\mu$ has been given by 
\begin{equation}
T_\mu = i\int e^{iqx}
{\langle{\pi_2(p')|T[A_\mu^{(\bar\pi_1)}H_w]|K(p)}\rangle}d^4x .
\end{equation}
$M_{\rm S}^{(n)}$ and $M_{\rm S}^{(l)}$ are pole amplitudes in the 
$s$- and $u$-channels, respectively, i.e., 
\begin{eqnarray} 
&& \hspace{-10mm}
M_{\rm S}^{(n)}(K\rightarrow \pi_1\pi_2)  \nonumber\\
&&
= {i \over \sqrt{2}f_{\pi}}
\Bigl({m_{\pi}^2 - m_{K}^2 \over m_n^2 - m_{K}^2}\Bigr)
  \langle{\pi_2|A_{\bar \pi_1}|n}\rangle
 \langle{n|H_w|K}\rangle  + (\pi_1 \leftrightarrow \pi_2), 
                                                \label{eq:surf-n}\\
&& \hspace{-10mm}
M_{\rm S}^{(l)}\,(K\rightarrow \pi_1\pi_2) \nonumber\\
&&= {i \over \sqrt{2}f_{\pi}}\Bigl({m_{\pi}^2 - m_{K}^2 
                              \over m_l^2 - m_{\pi}^2}\Bigr)
\langle{\pi_2|H_w|l}\rangle
                  \langle{l|A_{\bar \pi_1}|K}\rangle
+ (\pi_1 \leftrightarrow \pi_2).               \label{eq:surf-l}
\end{eqnarray}
In this way, an approximate $\sigma$ pole amplitude for the 
$K_S\rightarrow \pi^+\pi^-$ decay can be again described in terms of 
$\langle{\pi^-|A_{\pi^-}|\sigma}\rangle$ as 
\begin{equation}
M^{(\sigma)}(K_S\rightarrow \pi^+\pi^-) 
\simeq i{2 \over f_{\pi}}
\Bigl({m_{\pi}^2 - m_{K}^2 \over m_\sigma^2 - m_{K}^2}\Bigr)
  \langle{\pi^-|A_{\pi^-}|\sigma}\rangle
                 \langle{\sigma|H_w|K^0}\rangle.  \label{eq:K-sigma}
\end{equation}

Dominance of $\sigma$-meson pole in the $K_S\rightarrow\pi\pi$
amplitudes implies that $M^{(\sigma)}$ is much larger than the other 
contributions (the other pole amplitudes and $M_{\rm ETC}$ in 
addition to the factorized one, $M_{\rm fact}$, if it exists), i.e., 
\begin{equation}
|M^{(\sigma)}| \gg |M_{\rm ETC}|,\,\,|M_{\rm S}^{(n\neq\sigma)}|,
\,\, |M_{\rm S}^{(l)}|,\,\,|M_{\rm fact}|, 
\end{equation} 
unless the amplitudes in the right-hand-side cancel accidentally each
other. However, if the $\sigma$ pole contribution dominates 
$K_S\rightarrow\pi\pi$ amplitudes, it may be worried about that its 
strange partner, $\kappa$, also plays a role in the same amplitudes. 
The $\kappa$ pole amplitude can be obtained in the same way as 
$M^{(\sigma)}$ and its ratio to $M^{(\sigma)}$ is approximately given 
by 
\begin{equation}
\Big|{M^{(\kappa)}(K_S\rightarrow\pi^+\pi^-) 
\over M^{(\sigma)}(K_S\rightarrow\pi^+\pi^-)} \Big|
\sim
\Big|\Bigl({m_\sigma^2 -m_K^2 \over m_\kappa^2}\Bigr) 
{\langle{\pi|H_w|\kappa}\rangle 
\over \langle{\sigma|H_w|K}\rangle}\Big| . 
\end{equation}
If $m_\kappa^2 > m_\sigma^2 \sim m_K^2$, the above ratio will be 
small unless $\langle{\pi|H_w|\kappa}\rangle$ is anomalously
enhanced. However, if $m_\kappa^2 \sim m_\sigma^2 \gg m_K^2$, the 
$\kappa$ pole can play a role in the $K_S\rightarrow\pi\pi$ 
amplitudes. Nevertheless, neglect of $\kappa$ pole contribution does 
not change the essence of the physics in the $K_L$-$K_S$ mass 
difference as will be seen later. Therefore, we will neglect the 
$\kappa$ contribution to the $K_S\rightarrow\pi\pi$ amplitudes for 
simplicity. 

The decay rates for $\sigma\rightarrow\pi^+\pi^-$ and 
$K_S\rightarrow\sigma\rightarrow\pi^+\pi^-$ are given by 
\begin{equation}
\Gamma(\sigma\rightarrow\pi^+\pi^-)
\simeq {q_\sigma\over 4\pi f_\pi^2m_\sigma^2}
\bigl(m_\sigma^2 - m_\pi^2\bigr)^2
|\langle{\pi^-|A_{\pi^-}|\sigma}\rangle|^2,
                                          \label{eq:Gamma_sigma}
\end{equation}
and 
\begin{equation}
\Gamma^{(\sigma)}(K_{\rm S}\rightarrow\pi^+\pi^-)
\simeq {q_K\over 2\pi f_\pi^2m_K^2}
\Bigl({m_K^2 - m_\pi^2 \over m_K^2 - m_\sigma^2}\Bigr)^2
| \langle{\pi^-|A_{\pi^-}|\sigma}\rangle
                 \langle{\sigma|H_w|K^0}\rangle|^2,
                                        \label{eq:Gamma_K-sigma}
\end{equation}
respectively, where $q_\sigma$ and $q_K$ are the center-of-mass 
momenta of the final pions in the corresponding decays. Since the 
$K_S\rightarrow\pi\pi$ mode dominates the decays of $K_S$, its 
total width, $\Gamma_{K_S}$, is approximately given by 
$\Gamma_{K_S}\simeq {3\over 2}\Gamma(K_S\rightarrow \pi^+\pi^-)$,  
so that 
$\Gamma_{K_S} 
\simeq {3\over 2}\Gamma^{(\sigma)}(K_S\rightarrow \pi^+\pi^-)$  
under the $\sigma$ pole dominance hypothesis. 

The $\sigma$ meson pole dominance in the $K_S\rightarrow\pi\pi$ means 
that the matrix element, $\langle{\sigma|H_w|K}\rangle$, exists and 
its magnitude should be sizable. Therefore, under this hypothesis, 
the $\sigma$ meson pole may give a substantial contribution to 
$\Delta m_K$. The formula describing dynamical contributions of 
hadrons to $\Delta m_K$ has been given in the IMF long time 
ago~\cite{rf:Itzykson}. Using it, we obtain the following pole 
contribution of $\sigma$ meson, 
\begin{equation}
\Delta m_K^{(\sigma)}
= - {|\langle{K_L|H_w|\sigma}\rangle|^2 
\over 2m_K(M_K^2 - m_{\sigma}^2)},
                                           \label{eq:m_K-sigma}
\end{equation}
where the matrix element, $\langle{K_L|H_w|\sigma}\rangle$, is again 
evaluated in the IMF. For later convenience, we consider the ratio of
the $K_L$-$K_S$ mass difference to the full width of $K_S$. If we 
assume the $\sigma$ pole dominance in the $K_S\rightarrow\pi\pi$ 
decays, we obtain 
\begin{equation}
R^{(\sigma)} \equiv {\Delta m_K^{(\sigma)}\over \Gamma_{K_S}}
\simeq {1\over 2}\Bigl({q_\sigma \over q_K}\Bigr)
{(m_\sigma^2 - m_K^2)\over m_\sigma^2}
\Bigl({m_\sigma^2-m_\pi^2\over m_K^2-m_\pi^2}\Bigr)^2
{m_K\over \Gamma_\sigma}                      \label{eq:ratio}
\end{equation}
from Eqs.~(\ref{eq:Gamma_sigma}) -- (\ref{eq:m_K-sigma}), where 
the full width of $\sigma$ is given by 
$\Gamma_\sigma \simeq {3\over 2}\Gamma(\sigma\rightarrow\pi^+\pi^-)$  
for $m_\sigma$ less than the $K\bar K$ threshold ($\simeq 1$ GeV).  

Now we study whether the above $\sigma$ pole dominance in the 
$K_S\rightarrow\pi\pi$ decays can be realized in consistency with 
$\Delta m_K$. 
\begin{center}
\hspace{0mm}
\epsfxsize= 0.5\hsize
\epsffile{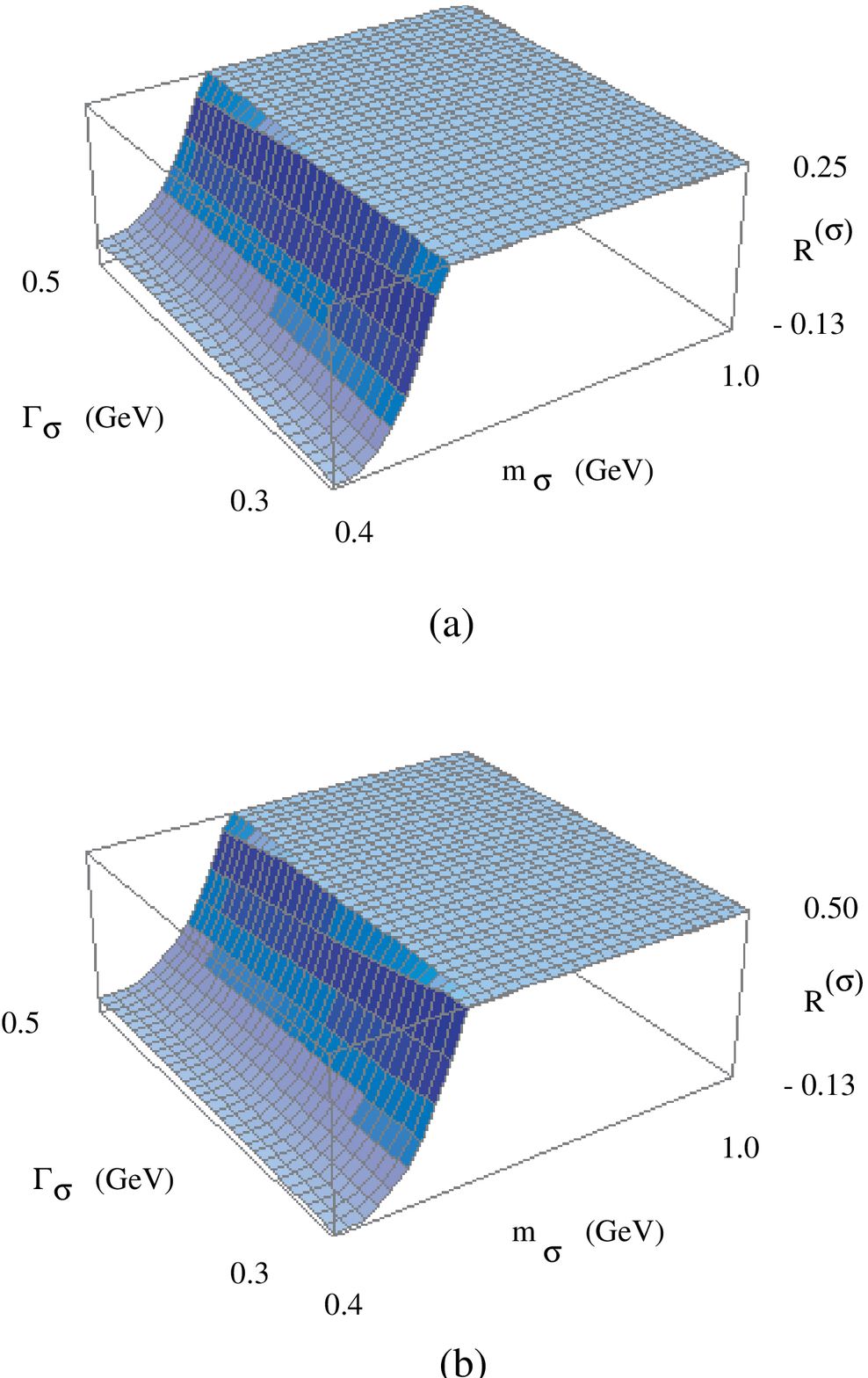}
\vspace{10mm}
\begin{minipage}{140mm}
{Fig.~I. 
$R^{(\sigma)}={\Delta m^{(\sigma)}/\Gamma_{K_S}}$ for 
$0.4 < m_\sigma < 1.0$ GeV and $0.3 < \Gamma_\sigma < 0.5$ GeV. 
$R^{(\sigma)}$ is cut at 0.25 in (a) and at 0.50 in (b) in order not 
to exceed the estimated $\pi\pi$ continuum contribution 
$R^{(\pi\pi)}$ and the measured $R_{\rm exp}$, respectively, as 
discussed in the text.
}
\end{minipage}
\end{center}
\vspace{10mm}
It has been known that contribution of $S$-wave $\pi\pi$ intermediate 
states to $\Delta m_K$ can occupy about a half~\cite{rf:Pennington} 
of the observed value~\cite{rf:PDG'98}, i.e., 
\begin{equation}
R^{(\pi\pi)} \equiv {\Delta m_K^{(\pi\pi)} \over \Gamma_{K_S}}
= 0.22 \pm 0.03, \qquad
R_{\rm exp} \equiv {\Delta m_K \over \Gamma_{K_S}}\Big|_{\rm exp} 
= 0.477 \pm 0.022. 
                                                \label{eq:m_K-data}
\end{equation}
The above $\Delta m_K^{(\pi\pi)}$ was estimated by using the 
Muskhelishvili-Omn\`es equation and the measured $\pi\pi$ phase 
shifts, etc., in which any indication of $\sigma$ meson was not
obviously seen. Therefore, if $\sigma$ exists, its contribution 
should be included in the above $\Delta m_K^{(\pi\pi)}$, so that 
we may put loosely the upper limit of the $\sigma$ pole 
contribution to $\Delta m_K$ around the above estimate of 
$\Delta m_K^{(\pi\pi)}$, i.e., 
${\Delta m_K^{(\sigma)}/ \Gamma_{K_S}} < 0.25$, 
and look for values of $m_\sigma$ and $\Gamma_\sigma$ to satisfy it 
since $\sigma$ meson is still hypothetical, i.e., its mass and width 
are still not confirmed. At energies lower than 900 MeV, the $\pi\pi$ 
phase shift analyses have excluded any narrow $I=0$ scalar state but 
a broad one ($\Gamma_\sigma \sim 500$ MeV) may have a room in the 
region~\cite{rf:PDG'98}, $0.4 < m_\sigma < 1.2$ GeV. In fact, various 
broad candidates of $\sigma$ meson with different masses 
($\sim 500 - 700$ MeV), different widths ($\sim 300 - 600$ MeV) and 
different structures have been studied at this 
workshop~\cite{rf:Tornqvist}.

$R^{(\sigma)}$ in Eq.(\ref{eq:ratio}) increases rapidly as $m_\sigma$ 
increases. It is beyond not only the estimated $R^{(\pi\pi)}$ for 
$m_\sigma > 0.55$ GeV but also the measured $R_{\rm exp}$  in 
Eq.(\ref{eq:m_K-data}) for $m_\sigma > 0.57$ GeV and is much larger 
than the above cuts in the region $m_\sigma^2 \gg m_K^2$. Therefore, 
even if $\kappa$ pole contribution to the $K\rightarrow\pi\pi$ decays 
is taken into account, the result, $R^{(\sigma)} \gg R_{\rm exp}$ for 
$m_\sigma^2 \gg m_K^2$, is not changed as discussed before. In this
way, it is seen that the $\sigma$ meson pole dominance in the 
$K\rightarrow\pi\pi$ amplitudes is not compatible with $\Delta m_K$ 
if $m_\sigma > 0.57$ GeV and $0.3 < \Gamma_\sigma < 0.5$ GeV, unless 
any other contribution cancels $\Delta m_K^{(\sigma)}$. 

However, the above does not necessarily imply that the $\sigma$ meson
pole dominance is compatible with the $K_L$-$K_S$ mass difference if 
$m_\sigma < 0.55$ GeV, since we have so far considered only the long 
distance effects on the $K_L$-$K_S$ mass difference. The short 
distance contribution from the box diagram~\cite{rf:Gaillard-Lee} 
which is estimated by using the factorization may saturate the 
observed $(\Delta m_K)_{\rm exp}$ although it is still ambiguous 
because of uncertainty of the so-called $B_K$ parameter. If it is the 
case, however, we need some other contribution to cancel the $\pi\pi$ 
continuum contribution (including $\sigma$ meson pole). Possible 
candidates are pseudo-scalar(PS)-meson poles since the other 
contributions of multi hadron intermediate states will be small 
because of their small phase space volumes. The above implies that 
the matrix elements,  
$\langle{P|H_w|K}\rangle,\,P=\pi^0,\,\eta,\,\eta',\,\cdots$, 
survive and their sizes are large enough to cancel 
$\Delta m_K^{(\pi\pi)}$. In this case, however, 
$\langle{\pi|H_w|K}\rangle$'s can give large effects on the 
$K\rightarrow\pi\pi$ amplitudes~\cite{rf:dalitz-decay} through 
Eq.(\ref{eq:hard pion}) with Eq.(\ref{eq:ETC}) and break the 
$\sigma$ meson pole dominance. 

For the $K_L\rightarrow\gamma\gamma$ decay, it is known that short 
distance contribution is small~\cite{rf:Gaillard-Lee}. To reproduce 
the observed rate for this decay, we again need contributions of 
PS-meson poles given by the matrix elements, 
$\langle{P|H_w|K}\rangle$'s, with sufficient magnitude, although 
their contributions are sensitive to the $\eta$-$\eta'$ mixing and 
are not always sufficient. In fact, the above PS-meson matrix 
elements can approximately reproduce 
$\Gamma(K_L\rightarrow\gamma\gamma)_{\rm exp}$, 
$\Gamma(K\rightarrow\pi\pi)_{\rm exp}$'s and 
$(\Delta m_K)_{\rm exp}$, 
simultaneously, with the help of some other contributions 
(non-factorizable amplitudes with PS- and $K^*$-meson poles for the 
$K_L\rightarrow\gamma\gamma$ decay, factorized ones for the 
$K\rightarrow\pi\pi$ decays and the short distance contribution to 
the $K^0$-$\bar K^0$ mixing, etc.) but without any contribution of 
$\sigma$ pole~\cite{rf:dalitz-decay}. Namely, we do not necessarily 
need the $\sigma$ pole contribution in the $K_S\rightarrow\pi\pi$ 
decays. 

As was seen above, it is unlikely that the $\sigma$ meson pole 
amplitude dominates the $K_S\rightarrow\pi\pi$. It will be seen
directly by comparing $M^{(\sigma)}(K_S\rightarrow \pi^+\pi^-)$ with 
$M_{\rm ETC}(K_S\rightarrow \pi^+\pi^-)$. If the asymptotic matrix
elements, $\langle{\pi|H_w|K}\rangle$'s, with sufficient magnitude 
exist and satisfy the $|\Delta {\bf I}|={1\over 2}$ rule (as derived 
by using a simple quark model~\cite{rf:dalitz-decay} or as required 
to realize the same rule in the $K\rightarrow\pi\pi$ amplitudes, i.e., 
$M_{\rm ETC}(K^+\, \rightarrow \pi^+\pi^0)=0$), we obtain  
\begin{equation}
\Big|{M^{(\sigma)}(K_S^0\, \rightarrow \pi^+\pi^-)
\over M_{\rm ETC}(K_S^0\, \rightarrow \pi^+\pi^-)}\Big| 
\simeq 2\Big|\Bigl({m_K^2 - m_\pi^2 \over m_\sigma^2 - m_K^2}\Bigr)
\langle{\pi^-|A_{\pi^-}|\sigma}\rangle
{\langle{\sigma|H_w|K^0}\rangle 
\over \langle{\pi^+|H_w|K^+}\rangle}\Big|.  
\end{equation}
The mass dependent factor $|(m_K^2-m_\pi^2)/(m_\sigma^2-m_K^2)|$ from 
$M^{(\sigma)}$ can be enhanced only if $m_\sigma$ is very close to 
$m_K$ and $\sigma$ is narrow. However, if $\Gamma_\sigma$ were small, 
$|\langle{\pi^-|A_{\pi^-}|\sigma}\rangle|$ also would be small.
When we smear out the singularity at $m_\sigma = m_K$ using the 
Breit-Wigner form, the size of 
$|{(m_K^2 - m_\pi^2)/(m_\sigma^2 - m_K^2)}
\langle{\pi^-|A_{\pi^-}|\sigma}\rangle|$
is at most $\simeq 2$ for 
$0.4 < m_\sigma < 1.0$ GeV and $0.3 < \Gamma_\sigma < 0.5$ GeV. 
However, any narrow $\sigma$ state around $m_K$ is not 
allowed~\cite{rf:PDG'98} as mentioned before. Moreover, $\sigma$ does 
not belong to the same ground state as $\pi$ and $K$ (for example, 
$^3P_0$ of $\{q\bar q\}$ state in the quark model, etc.), so that  
the matrix elements, $|\langle{\sigma|H_w|K}\rangle|$, will 
be much smaller than $|\langle{\pi|H_w|K}\rangle|$ since wave 
function overlapping between $\sigma$ and $K$ meson states will be 
much smaller than that between $\pi$ and $K$ which belong to the same 
$^1S_0$ state of $\{q\bar q\}$. Therefore, it is unlikely that the 
$\sigma$ meson pole amplitude dominates the $K\rightarrow\pi\pi$ 
amplitudes. 

An amplitude for dynamical hadronic process can be decomposed into 
({\it continuum contribution}) + ({\it  Born term}).  
Since $M_{\rm S}$ has been given by a sum of pole amplitudes, 
$M_{\rm ETC}$ corresponds to the continuum 
contribution~\cite{rf:MATHUR}. In the present case, 
$M_{\rm ETC}(K_S\rightarrow\pi\pi)$ will be dominated by 
contributions of isoscalar $S$-wave $\pi\pi$ intermediate states and 
develop a phase ($\simeq$ isoscalar $S$-wave $\pi\pi$ phase shift at 
$m_K$) relative to the Born term which is usually taken to be real in 
the narrow width limit. The estimated phase difference between 
$|\Delta{\bf I}|={1\over 2}$ and ${3\over 2}$ amplitudes for the
$K\rightarrow\pi\pi$ decays is close to the measured isoscalar 
$S$-wave $\pi\pi$ phase shift at $m_K$~\cite{rf:Kamal}. It suggests  
that the isoscalar $S$-wave $\pi\pi$ continuum contribution will be 
dominant in the $K_S\rightarrow\pi\pi$ amplitudes. 

In summary, we have studied contribution of the $\sigma$ meson pole 
to $\Delta m_K$ under the hypothesis that $\sigma$ meson pole 
dominates the $K_S\rightarrow\pi\pi$ amplitudes, and have seen that 
it provides too large contributions to $\Delta m_K$ and that, to
cancel out such effects, contributions of pseudo-scalar-meson poles 
will be needed. We also have discussed, comparing the $\sigma$ meson 
pole amplitude with $M_{\rm ETC}$ in the $K_S\rightarrow\pi\pi$ 
amplitudes, that enhancement of the $\sigma$ meson pole contribution 
is not sufficient if it is broad. Additionally, a recent analysis in 
the $K\rightarrow\pi\pi$ decays within the theoretical framework of 
non-linear $\sigma$ model suggests that the $\sigma$ meson pole 
contribution can occupy, at most, about a half of the 
$|\Delta{\bf I}|={1\over 2}$ amplitude~\cite{rf:Harada}. Therefore, 
we conclude that the $\sigma$ pole dominance in the 
$|\Delta{\bf I}|={1\over 2}$ amplitude for the $K\rightarrow\pi\pi$ 
decays is very unlikely. 

\vspace{0.5cm}

The author thanks Dr.~D.-X.~Zhang and Dr.~Y.~Y.~Keum for discussions
and comments. This work was supported in part by the Grant-in-Aid 
from the JSPS. 


\end{document}